\documentclass[preprint]{aastex}
\usepackage{natbib}
\usepackage{booktabs}
\usepackage{threeparttable}
\usepackage{CJK}

\newcommand{\fexxi}{Fe \scriptsize{XXI} \normalsize}
\newcommand{\ci}{C \scriptsize{I} \normalsize}
\newcommand{\oi}{O \scriptsize{I} \normalsize}
\newcommand{\siiv}{Si \scriptsize{IV} \normalsize}
\newcommand{\feii}{Fe \scriptsize{II} \normalsize}
\newcommand{\siii}{Si \scriptsize{II} \normalsize}
\newcommand{\si}{S \scriptsize{I} \normalsize}

\begin{document}

\title{Observations of Electron-driven Evaporation during a Flare Precursor}

\author{Dong Li\altaffilmark{1,2,3}, Ying Li\altaffilmark{1}, Wei Su\altaffilmark{4}, Yu Huang\altaffilmark{1},
         and Zongjun Ning\altaffilmark{1}}
 \affil{$^1$Key Laboratory of Dark Matter and
Space Astronomy, Purple Mountain Observatory, CAS, Nanjing 210008,
China} \affil{$^2$CAS Key Laboratory of Solar Activity, National
Astronomical Observatories, Beijing 100012, China} \affil{$^3$Sate
Key Laboratory of Space Weather, Chinese Academy of Sciences,
Beijing 100190, China} \affil{$^4$ MOE Key Laboratory of Fundamental
Physical Quantities Measurements, School of Physics, Huazhong
University of Science and Technology, Wuhan, 430074, China}
\altaffiltext{}{Correspondence should be sent to: lidong@pmo.ac.cn,
ningzongjun@pmo.ac.cn}

\begin{abstract}
We investigate the relationship between the blue shifts of a hot
emission line and the nonthermal emissions in microwave and hard
X-ray (HXR) wavelengths during the precursor of a solar flare on
2014 October 27. The flare precursor is identified as a small but
well-developed peak in soft X-ray and extreme-ultraviolet passbands
before the {\it GOES} flare onset, which is accompanied by a
pronounced burst in microwave 17~$\&$~34~GHz and HXR 25$-$50~keV.
The slit of {\it Interface Region Imaging Spectrograph} ({\it IRIS})
stays at one ribbon-like transient during the flare precursor, where
shows visible nonthermal emissions in NoRH and {\it RHESSI} images.
The {\it IRIS} spectroscopic observations show that the hot line of
\fexxi~1354.09~{\AA} (log$T\sim$~7.05) displays blue shifts, while
the cool line of \siiv~1402.77~{\AA} (log$T\sim$~4.8) exhibits red
shifts. The blue shifts and red shifts are well correlated to each
other, indicative of an explosive chromospheric evaporation during
the flare precursor particularly combining with a high nonthermal
energy flux and a short characteristic timescale. In addition, the
blue shifts of \fexxi 1354.09~{\AA} are well correlated with the
microwave and HXR emissions, implying that the explosive
chromospheric evaporation during the flare precursor is driven by
nonthermal electrons.
\end{abstract}


\section{Introduction}
Chromospheric evaporation is a well-known process during a solar
flare, which was first described by \cite{Neupert68}. This process
occurs when the chromospheric materials are heated more quickly than
they can radiatively cool \citep{Fisher85}. The heated materials
expand rapidly upward into the corona with a low density along the
reconnected magnetic field lines, and then those hot plasmas fill up
the newly formed flare loops which can be seen in soft X-ray (SXR)
or extreme-ultraviolet (EUV) passbands
\citep{Liu06,Ning10,Milligan15,Li17}. Usually, the emission lines
formed at a high temperature show blue shifts, which provides a
strong evidence for chromospheric evaporation
\citep{Czaykowska99,Liu09,Milligan09,Brosius16,Lee17}. Due to
momentum balance, some materials move slowly downward into the
chromosphere with a high density, which is supported by the red
shifts in the emission lines formed at a low temperature
\citep{Teriaca06,Tian15,Zhang16a,Li17b}. Notice that the red shifts
in cool emission lines might be not observed in some chromospheric
evaporations \citep{Milligan06,Brosius09,Raftery09}. In the
chromospheric regions with a high density, the energy is lost
through coulomb collision between the precipitating electrons and
the ambient plasmas, producing the hard X-ray (HXR) or microwave
emissions \citep{Brown71,Asai06}.

Chromospheric evaporation can be detected in multi-wavelengths,
ranging from HXR \citep{Liu06,Ning11,Zhang13} through EUV/UV
\citep{Czaykowska99,Li15,Tian18} to microwave
\citep{Aschwanden95,Ning09} channels. In HXR or EUV imaging
observations, the material movement from double footpoints to loop
top along the flare loops is considered to be the HXR/EUV signature
of chromospheric evaporation \citep{Ning10,Ning11b,Zhang13,Li17}. On
the dynamic spectra, the microwave emission is suddenly cut off in
the higher frequency and drifts to the lower frequency, which is
believed as the radio signature of chromospheric evaporation
\citep{Aschwanden95,Karlicky98,Ning09}. In spectroscopic
observations, Doppler shifts in the emissions lines formed at
different temperatures are often used to study chromospheric
evaporation. The speeds of hot lines formed in the corona are
observed as fast as around 100$-$400~km~s$^{-1}$, while the speeds
of cool lines formed in the chromosphere or transition region are
only about 10$-$50~km~s$^{-1}$
\citep{Ding96,Veronig10,Tian14,Brosius16}. This is because the
plasma density in the underlying chromosphere or transition region
is much larger than that in the overlying corona on the Sun
\citep{Fisher85,Doschek13,Milligan15}.

Chromospheric evaporation proceeds ``explosively" when the input
energy flux exceeds a critical value of
$\sim$10$^{10}$~erg~s$^{-1}$~cm$^{-2}$
\citep{Fisher85,Fisher85b,Zarro88,Kleint16}. The hot lines in the
corona appear blue shifts, while the cool lines in the chromosphere
or transition region appear red shifts
\citep{Feldman80,Del06,Brosius10,Chen10,Brosius15,Tian15,Lee17}.
Meanwhile, Chromospheric evaporation proceeds ``gently" if the input
energy flux is less than the critical value, and all the emission
lines from chromosphere through transition region to corona appear
blue shifts \citep{Milligan06,Brosius09,Raftery09,Li11,Sadykov15}.
It should be mentioned that the critical value of energy flux
between the ``explosive" and ``gentle" chromospheric evaporation
depends also on the other beam parameters \citep{Fisher89}, duration
of heating \citep{Reep15}, and acceleration nature \citep{Rubio15}.
For example, the explosive evaporation can be driven by stochastic
acceleration even with the very low energy flux. Until now, three
mechanisms have been proposed to explain chromospheric evaporation.
The first one emphases that the nonthermal electron-beams
accelerated by magnetic reconnection play a key role in driving
chromospheric evaporation \citep[or electron-driven,
e.g.,][]{Brosius03,Milligan09,Tian14,Tian15}. The second one is
thermal conduction, which states that the thermal energy can drive
chromospheric evaporation directly \citep{Fisher85,Falewicz09}. The
last one is the dissipation of Alfv\'{e}n waves \citep{Reep16}.

Chromospheric evaporation usually occurs during the impulsive phase
of a solar flare \citep{Brosius04,Brosius07,Brosius10,Tian15,Li17},
as stated in the standard flare model
\citep{Carmichael64,Sturrock66,Hirayama74,Kopp76}. It can also
happen in the decay or gradual phase of a solar flare
\citep{Zarro88,Czaykowska99,Li12}. However, chromospheric
evaporation in the pre-flare phase is relatively rarely reported. In
fact, before the {\it GOES} flare onset, the SXR light curve has
started to rise slowly or even shown a small but well-developed
peak, which is called flare precursor
\citep[e.g.,][]{Bamba13,Cheng16a,Li17c,Benz17,Shen17}. Sometimes,
the flare precursor can be identified as the chromospheric
brightening in EUV/UV images, which is thought to be related to the
characteristic structure of magnetic field \citep{Bamba13,Bamba17}.
Imaging and spectroscopic observations also show that there could
appear various precursors during the pre-flare phase, such as the
eruption and oscillation of magnetic flux rope
\citep{Cheng15,Cheng16a,Li16b,Zhou16,Li17c,Yan17}, the coronal
dimmings \citep{Zhang17}, and the upflows in active regions
\citep{Imada14,Dudik16,Woods17}, which suggest that the flare
precursors may play an important role in triggering solar flare. In
this paper, we detect chromospheric evaporation manifested by blue
shifts of the hot \fexxi~1354.09~{\AA} line and the cool
\siiv~1402.77~{\AA} line during the flare precursor that exhibits a
SXR/EUV peak. We also find a good correlation with high coefficients
of about 0.87$-$0.97 between the blue shifts and the microwave/HXR
emissions that show up before the {\it GOES} flare onset.

\section{Observations and Data Analysis}
\begin{figure}
\epsscale{1.0} \plotone{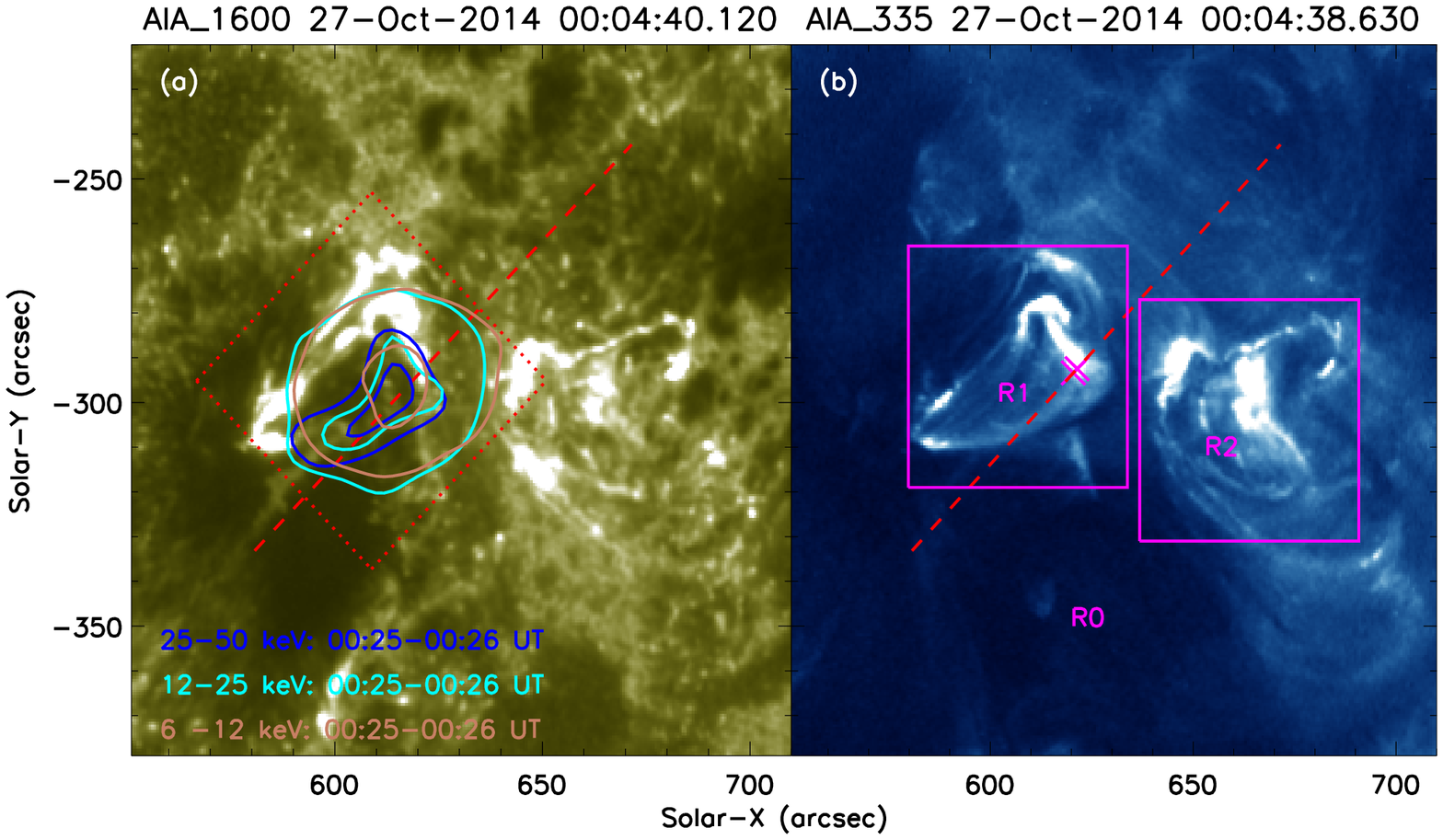} \caption{Snapshots of {\it SDO}/AIA
images in 1600~{\AA} (a) and 335~{\AA} (b). The contours are from
the {\it RHESSI} images (60\% $\&$ 90\%) in 6$-$12~keV (brown),
12$-$25~keV (turquoise), and 25$-$50~keV (blue) during the flare
peak time. The red dashed line represents the slit of {\it IRIS},
two purple short lines outline the studied locations, and the red
dotted diamond mark the region in Figure~\ref{image}. The purple
boxes mark the regions used to integrate the EUV fluxes in
Figure~\ref{flux}. \label{snap}}
\end{figure}

Our observations focus on the active region of NOAA AR12192 on 2014
October 27 between 00:01~UT and 00:06~UT. This active region is
simultaneously observed by the {\it Interface Region Imaging
Spectrograph} \citep[{\it IRIS},][]{Dep14}, the Atmospheric Imaging
Assembly \citep[AIA,][]{Lemen12} and Helioseismic and Magnetic
Imager \citep[HMI,][]{Schou12} aboard the {\it Solar Dynamics
Observatory} ({\it SDO}), the Nobeyama Radioheliograph
\citep[NoRH,][]{Hanaoka94}, and the {\it Reuven Ramaty High Energy
Solar Spectroscopic Imager} \citep[{\it RHESSI},][]{Lin02}.
Figure~\ref{snap} shows the snapshots of this active region in AIA
1600~{\AA} (a) and 335~{\AA} (b), respectively. The contours are
integrated from the {\it RHESSI} observations (i.e., detectors 3, 4,
5, 6, and 8) with the CLEAN algorithm between 00:25~UT and 00:26~UT
in 6$-$12~keV (brown), 12$-$25~keV (turquoise), and 25$-$50~keV
(blue), respectively. The red dashed line outlines the slit of {\it
IRIS}, which is along a 45$^{\circ}$ to the north$-$south direction.
Two purple short lines mark the location studied in this paper,
which is contained in the purple box of `R1'. While `R2' outlines
another EUV/UV bright region, and `R0' refers to the entire active
region in panel~(b).

\begin{figure}
\epsscale{1.0} \plotone{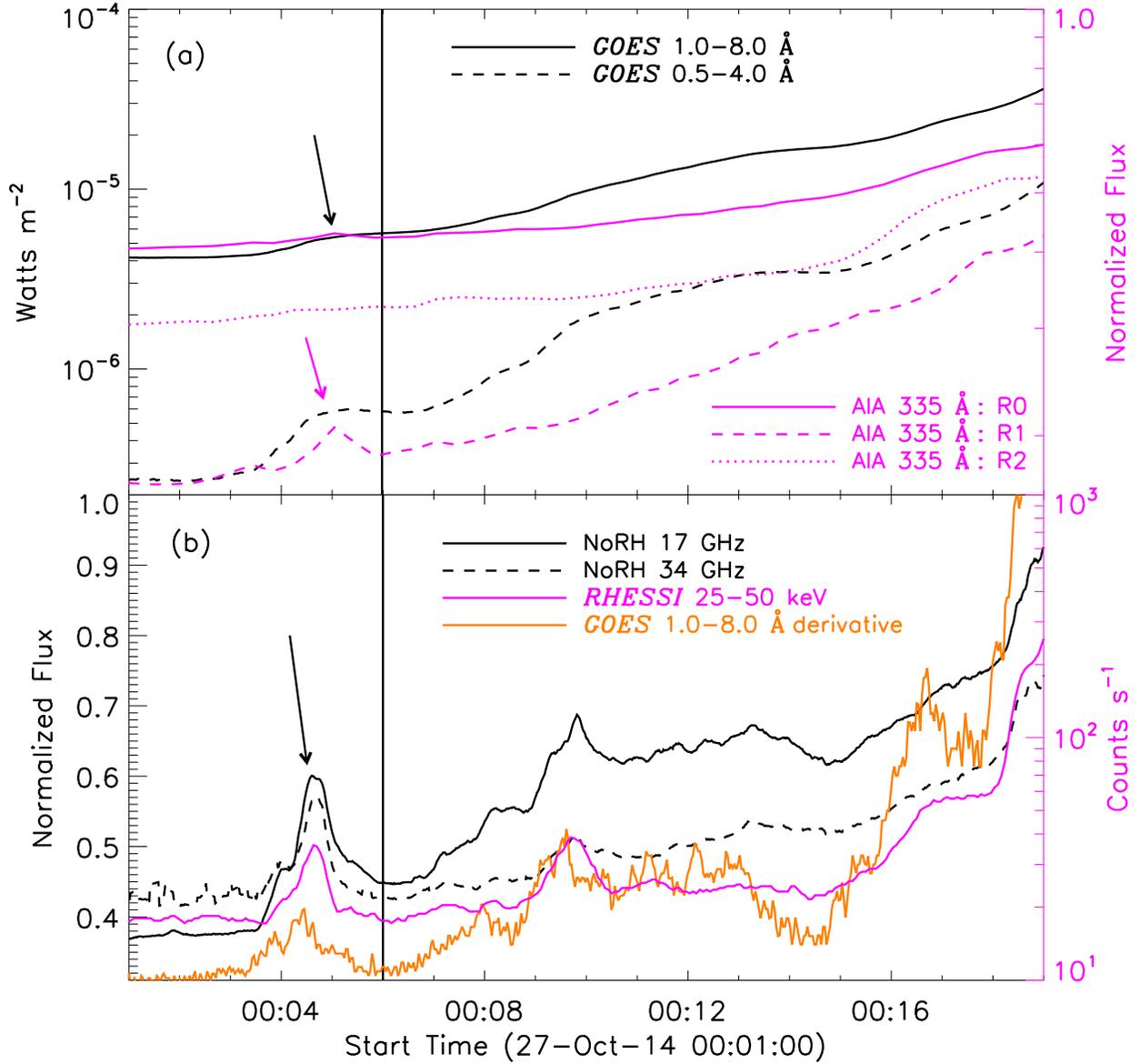} \caption{Panel~(a): {\it GOES} SXR
light curves from 00:01~UT to 00:19~UT on 2014 October 27. The
purple profiles are the normalized fluxes in AIA 335~{\AA} from
different regions (see Figure~\ref{snap}) on the Sun. The solid
vertical line marks the onset time of solar flare. Panel~(b):
Normalized fluxes between 00:01~UT$-$00:19~UT in NoRH microwave£¬
{\it RHESSI} HXR and {\it GOES} derivative channels. \label{flux}}
\end{figure}

Figure~\ref{flux}~(a) shows the SXR light curves in {\it GOES}
1.0$-$8.0~{\AA} (black solid) and 0.5$-$4.0~{\AA} (black dashed)
from 00:01~UT to 00:19~UT. A {\it GOES} M7.1 flare begins to burst
out at 00:06~UT (i.e., flare onset), as indicated by the dashed
vertical line. Prior to the {\it GOES} flare onset, a small SXR peak
appears at around 00:04:40~UT. It is much more pronounced in {\it
GOES} 0.5$-$4.0~{\AA} than that in {\it GOES} 1.0$-$8.0~{\AA}, as
indicated by the purple and black arrows. Notice that the {\it GOES}
SXR fluxes come from the full solar disk. To determine if the SXR
peak is related to the M7.1 flare, we use the spatially resolved
{\it SDO}/AIA observations. Therefore, the EUV fluxes (purple lines
in panel~(a)) in AIA~335~{\AA} are plotted, which are integrated
from the regions of R0 (purple solid), R1 (purple dashed), and R2
(purple dotted). Similar to the SXR light curves in {\it GOES}
1.0$-$8.0~{\AA} and 0.5$-$4.0~{\AA}, the EUV fluxes from the entire
active region (R0) reveal a faint peak, and in particular, the EUV
fluxes from the region of interest (R1) show a pronounced peak also
at around 00:04:40 UT. While the EUV fluxes from the other region
(R2) do not exhibit a corresponding peak at that time. This gives
the observational evidence that the EUV/SXR peak emission before
{\it GOES} flare onset is mainly from the studied locations (R1) and
related to our flare event. Moreover, the region R1 should be also
the main flaring region, which is indicated by the {\it RHESSI}
emissions from 00:25$-$00:26 UT around the flare peak time (see the
contours in Figure~\ref{snap}~(a)). All these observational facts
suggest that the SXR/EUV peak can be considered as the flare
percussor. Figure~\ref{flux}~(b) gives the normalized fluxes between
00:01~UT$-$00:19~UT in nonthermal emissions from NoRH 17~GHz (black
solid) and 34~GHz (black dashed), {\it RHESSI} 12$-$25~keV (purple),
and also {\it GOES} 1.0$-$8.0 derivative (orange). Same as the light
curves in SXR 0.5$-$4.0~{\AA} and EUV~335~{\AA}, both the microwave
and HXR fluxes exhibit a pronounced burst before the {\it GOES}
flare onset, i.e., from $\sim$00:03:30~UT to $\sim$00:05:50~UT, as
indicated by the black arrow.

\begin{figure}
\epsscale{1.0} \plotone{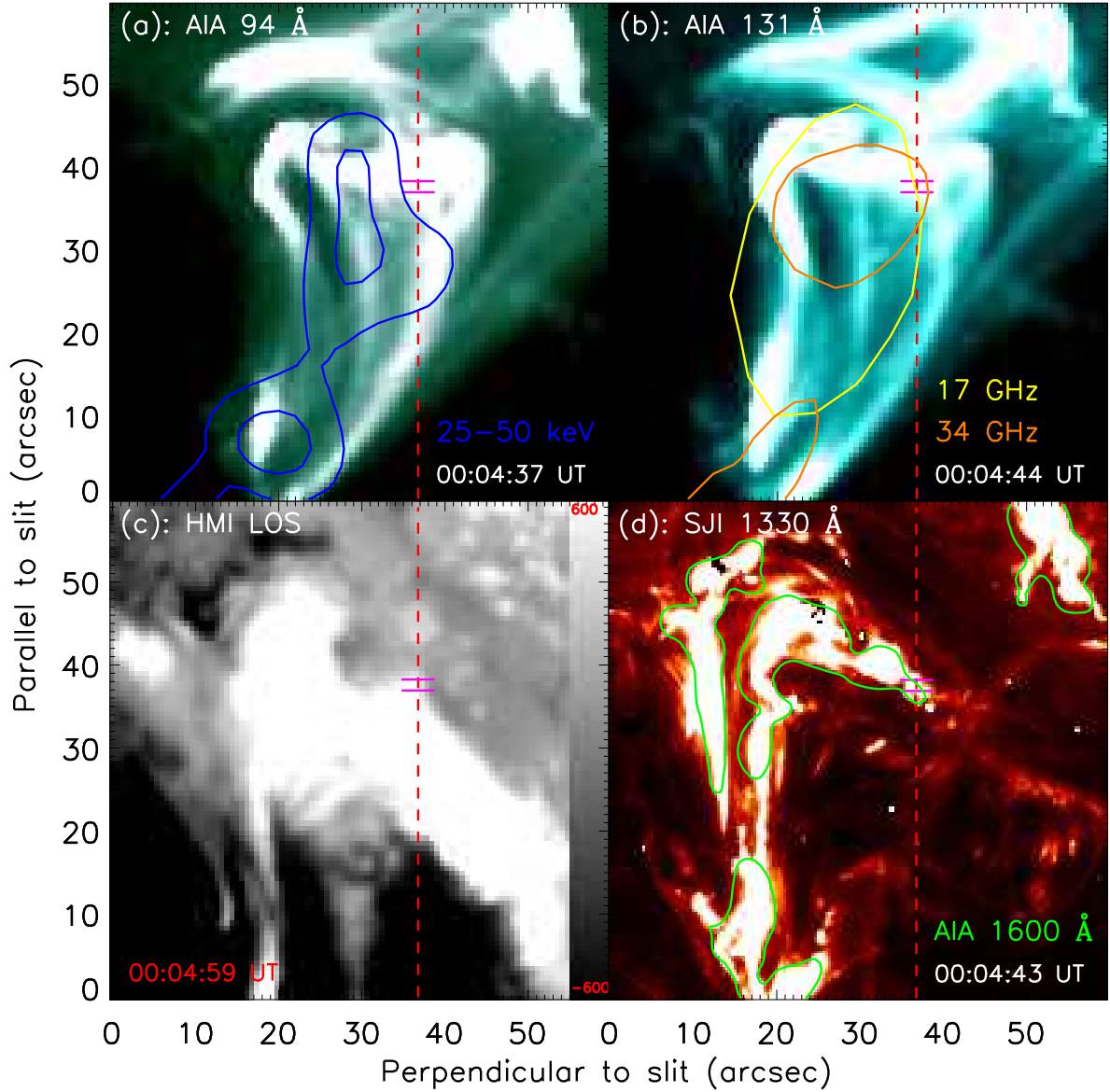} \caption{Multi-wavelength images
along {\it IRIS} slit direction around the SXR precursor peak of
solar flare. The blue contours represent the HXR emissions from {\it
RHESSI}, the levels are set at 70\% and 90\%. The yellow and orange
contours are the microwave emissions from NoRH. The green contours
represent the AIA 1600~{\AA} intensities. The red dashed line
outlines the slit of {\it IRIS}, and two purple short lines mark the
locations studied here. \label{image}}
\end{figure}

{\it IRIS} performs this observation in `sit-and-stare' mode from
18:52:50~UT on October 26 to 08:23:08~UT on October 27 in 2014,
covering the flare precursor. It points at a fixed center position
of (608\arcsec, -287\arcsec) with a max field-of-view (FOV) of
120\arcsec$\times$119\arcsec, which overlays with the region of R1.
Figure~\ref{image} shows the multi-wavelength images from {\it
SDO}/AIA, {\it SDO}/HMI, and {\it IRIS}/SJI at the peak time of
flare precursor. Here, X-axis is perpendicular to the slit of {\it
IRIS}, and Y-axis is parallel to the slit of {\it IRIS}. These
images have the same FOV of 60\arcsec$\times$60\arcsec, as marked by
the red dotted diamond in Figure~\ref{snap}~(a). Panels~(a) and (b)
show the intensity images in AIA 94~{\AA} and 131~{\AA}. The
overlaid contours represent HXR emission in 25$-$50~keV (blue
contours), microwave emissions in the frequencies of 17~GHz (yellow
contour) and 34~GHz (orange contours), respectively, which are
integrated from 00:04~UT to 00:05~UT. Panel~(c) displays the
line-of-sight (LOS) magnetogram. Panel~(d) gives the SJI 1330~{\AA}
image with the overplotted green contours taken from the AIA
1600~{\AA} intensities, which are applied to co-align with the SJI
1330~{\AA} image by cross-correlating \citep{Cheng15,Tian16}. We can
see that the bright features from these two passbands match well,
since they both contain the UV continuum emissions in the
temperature-minimum region. Figure~\ref{image} indicates that two
ribbon-like transients in SJI 1330~{\AA} (panel~d) are connected by
some hot coronal loops visible in AIA 94~{\AA} and 131~{\AA} images
(panel~a and b), which are rooted in the positive and negative
magnetic fields, respectively (panel~c). One of the ribbon-like
transients is crossed by the slit of {\it IRIS}, as marked by the
dashed vertical line. This location is also co-spatial with the
nonthermal source, such as HXR emission in 25$-$50~keV, microwave
emissions in the frequencies of 17~GHz and 34~GHz.

The hot emission line of \fexxi~1354.09 {\AA} and the cool emission
line of \siiv~1402.77~{\AA} have been used in many spectroscopic
studies to investigate chromospheric evaporation
\citep[e.g.,][]{Tian14,Li15b,Tian15,Brosius16,Zhang16a,Zhang16b,Li17,Li17b}.
It is well accepted that the forbidden line of \fexxi~1354.09 {\AA}
is a hot (log~$T\sim$~7.05) and broad emission line during solar
flares \citep{Doschek75,Cheng79,Mason86,Innes03a,Innes03b}.
Meanwhile, {\it IRIS} spectroscopic observations show that
\fexxi~1354.09~{\AA} is always blended with a number of cool and
narrow emission lines, which are from the neutral or singly ionized
species. Those blended emission lines can be easily detected in the
position of flare ribbon, including known and unknown emission
lines, such as, the \ci line at 1354.29~{\AA}, the \feii lines at
1353.02~{\AA}, 1354.01~{\AA}, and 1354.75~{\AA}, the \siii lines at
1352.64~{\AA} and 1353.72~{\AA}, the unidentified lines at
1353.32~{\AA} and 1353.39~{\AA}
\citep[e.g.,][]{Li15,Li16,Polito15,Polito16,Tian15,Tian16,Tian17,Young15}.
In order to extract the hot line of \fexxi~1354.09~{\AA} and the
cool line of \ci~1354.29~{\AA} \citep[log~$T\sim$~4.0,][]{Huang14},
we apply a multi-Gaussian functions superimposed on a linear
background to fit the {\it IRIS} spectrum at `\oi' window
\citep[e.g.,][]{Li15,Li16}, which has been pre-processed (i.e., {\it
IRIS} spectral image deformation, bad pixel despiking and wavelength
calibration) with the standard routines in Solar Soft Ware
\cite[SSW,][]{Freeland00}. In a word, the line positions and widths
of these blended emission lines are fixed or constrained, and their
peak intensities are tied to the isolated emission lines from the
similar species. More details can be found in our previous papers
\citep{Li15,Li16}. On the other hand, the cool line of
\siiv~1402.77~{\AA} (log~$T\sim$~4.8) at `\siiv' window is
relatively isolated, and it can be well fitted with a
single-Gaussian function superimposed on a linear background
\citep{Li14,Li17}. Using the relatively strong neutral lines (i.e.,
`\oi' 1355.60~{\AA} and `\si' 1401.51~{\AA}), we also perform an
absolute wavelength calibration for the spectra at `\oi' and `\siiv'
windows, respectively \citep{Tian15,Tian17}. Finally, the Doppler
velocities of \fexxi~1354.09~{\AA}, \ci~1354.29~{\AA}, and
\siiv~1402.77~{\AA} are determined by the fitting line centers
removing from their rest wavelengths, respectively
\citep{Cheng16b,Guo17,Li17}. As the hot \fexxi line is absent in the
non-flaring spectrum, the rest wavelength for \fexxi line (i.e.,
1354.09~{\AA}) is determined by averaging the line centers of the Fe
XXI profiles which used in the pervious {\it IRIS} observations
\citep{Brosius15,Brosius16,Polito15,Polito16,Sadykov15,Tian15,Young15,Lee17}.
While the rest wavelengths for \ci and \siiv lines, i.e.,
1354.29~{\AA} and 1402.77~{\AA}, respectively, are determined from
their quiet-Sun spectra \citep{Li14,Li15}.

\section{Results}
\begin{figure}
\epsscale{1.0} \plotone{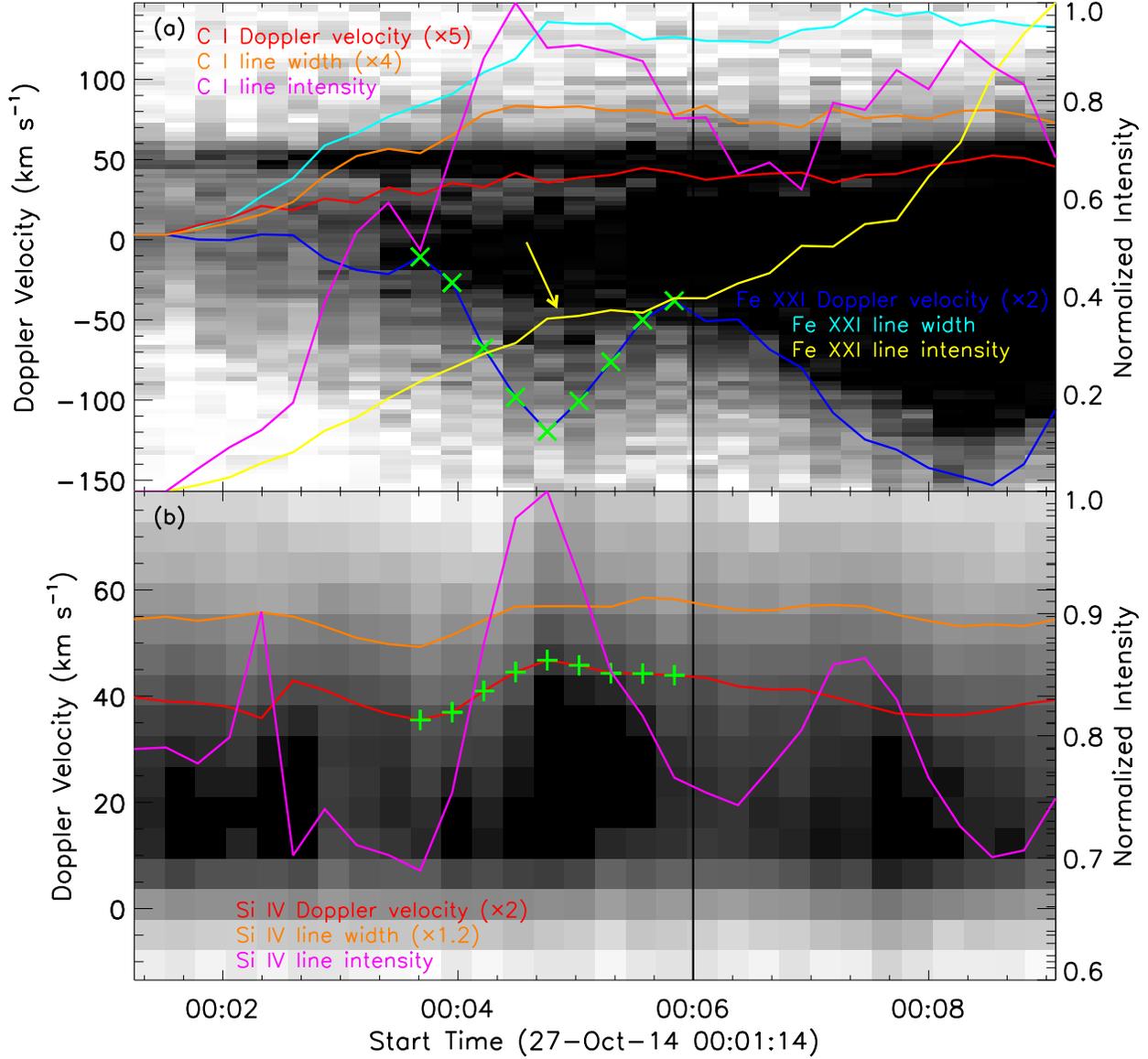} \caption{Time evolutions of the line
profiles at the windows of `\oi' (a) and `\siiv' (b) on the slit of
{\it IRIS}, and the zero velocity is set to the rest wavelength of
\fexxi~1354.09~{\AA} (a) or \siiv~1402.77~{\AA} (b). The overplotted
profiles are the time series of Doppler velocity, line intensity,
and line width in \fexxi~1354.09~{\AA}, \ci~1354.29~{\AA} and
\siiv~1402.77~{\AA}, respectively. The green signs mark the points
during the flare precursor, and the solid vertical line indicates
the onset time of solar flare. \label{vel}}
\end{figure}

Figure~\ref{vel} shows the time evolutions of the line profiles from
00:01:14~UT to 00:09:04~UT at the {\it IRIS} windows of `\oi' (a)
and `\siiv' (b), and the zero velocity is set to the rest
wavelengths of \fexxi~1354.09~{\AA} or \siiv~1402.77~{\AA},
respectively. They are averaged on the positions between
$\sim$36.9\arcsec\ and $\sim$38.3\arcsec\ along the slit of {\it
IRIS}, as marked by the two purple short lines in
Figure~\ref{image}. The overplotted lines in panel~(a) are the time
series of Doppler velocity (blue/red), line width (turquoise/orange)
and line intensity (yellow/purple) in \fexxi~1354.09~{\AA} and
\ci~1354.29~{\AA}, respectively, and the overplotted lines in
panel~(b) are the time series of Doppler velocity (red), line width
(orange) and line intensity (purple) in \siiv~1402.77~{\AA}. To
exhibit the time series clearly, we have multiplied a factor for
some time series. The Doppler velocities of \fexxi~1354.09~{\AA}
start to appear a precursor burst at $\sim$00:03:30~UT in the
blue-shifted wings, and peak at $\sim$00:04:40~UT, while disappear
before the {\it GOES} flare onset, which is $\sim$00:05:50~UT, as
shown by the green crosses (`$\times$'). During the same time
intervals, the Doppler velocities of \siiv~1402.77~{\AA} also show a
precursor burst but in the red-shifted wings, as indicated by the
green pluses (`+'). We note that the Doppler velocities of
\ci~1354.29~{\AA} do not exhibit such pronounced precursor burst,
but appear much more constant and flat in the red-shifted wings. The
hot line of \fexxi~1354.09~{\AA} exhibits blue shifts, and both the
cool lines of \ci~1354.29~{\AA} and \siiv~1402.77~{\AA} show red
shifts, indicating an explosive chromospheric evaporation before the
{\it GOES} flare oneset. The maximum speed of blue shifts in
\fexxi~1354.09~{\AA} during the flare precursor is about
60~km~s$^{-1}$, and the maximum speeds of red shifts in
\siiv~1402.77~{\AA} and \ci~1354.29~{\AA} during the flare precursor
are around 24~km~s$^{-1}$ and 8~km~s$^{-1}$, respectively.

During the flare precursor, the line widths of both hot
(\fexxi~1354.09~{\AA}) and cool (\ci~1354.29~{\AA} $\&$
\siiv~1402.77~{\AA}) emission lines demonstrate a small precursor
peak. This fact is mostly likely to reveal an energy release process
during this explosive chromospheric evaporation, which is used to
heat the local plasma. The enhancements of these line widths in both
hot and cool emission lines might also be caused by the nonthermal
broadening during the flare precursor. On the other hand, both the
line intensities in \ci~1354.29~{\AA} and \siiv~1402.77~{\AA} show a
pronounced precursor peak before the {\it GOES} flare onset.
Meanwhile, the line intensity of \fexxi~1354.09~{\AA} also exhibits
a faint precursor peak before the {\it GOES} flare onset, as marked
by the yellow arrow. Notice that the precursor peak in FUV emission
lines from the {\it IRIS} spectroscopic observations appears to well
agree with the EUV precursor peak in AIA~335~{\AA} from the imaging
observations (Figure~\ref{flux}).

\begin{figure}
\epsscale{1.0} \plotone{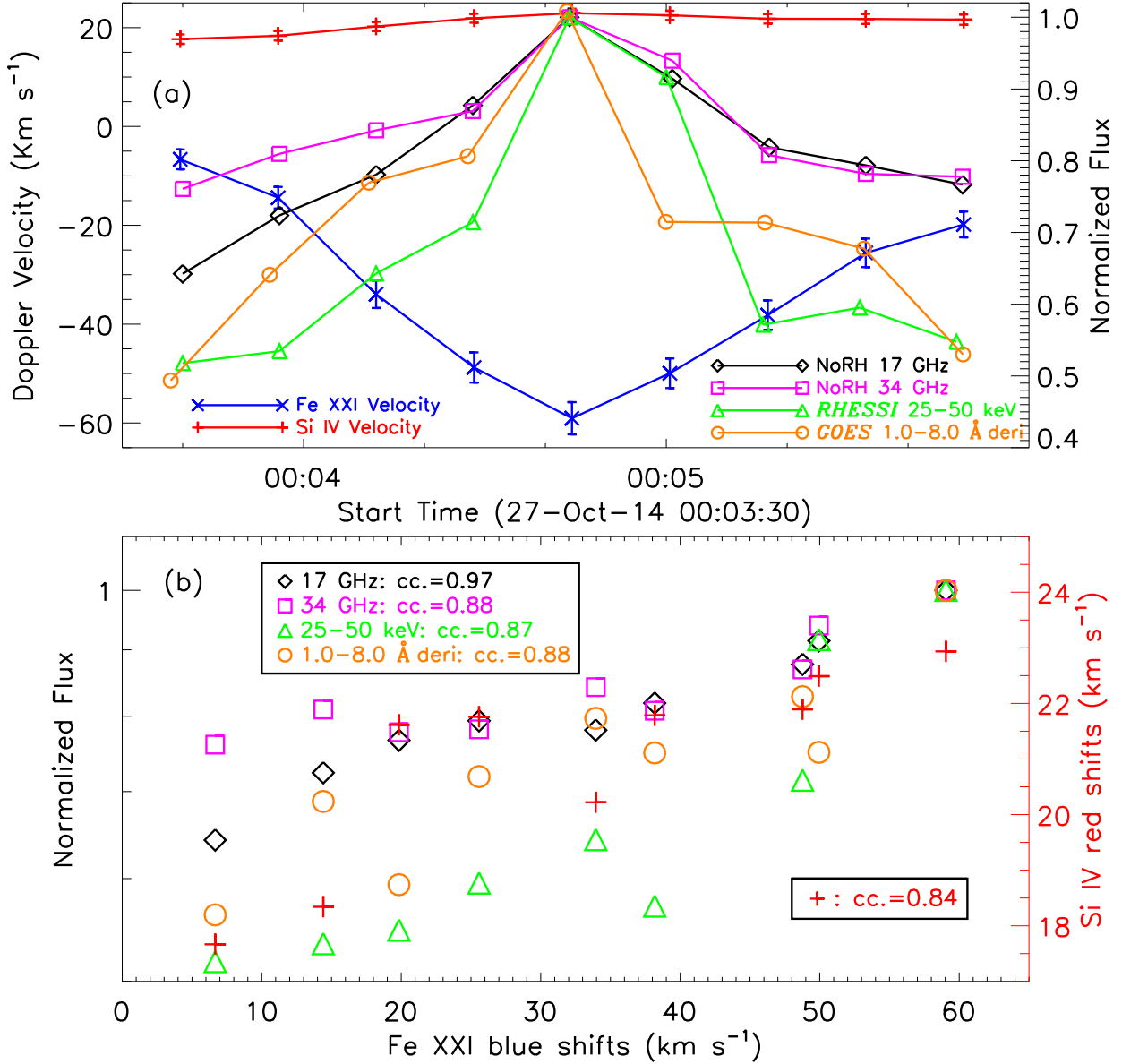} \caption{Panel~(a): Time series of
Doppler velocity with error bars in \fexxi~1354.09~{\AA} (blue) and
\siiv~1402.77~{\AA} (red), microwave fluxes in 17~GHz (black) and
34~GHz (purple), HXR light curves in 25$-$50~keV (green) and
1.0$-$8.0~{\AA} derivative (orange). All these time series are
forced to the same time cadence. Panel~(b): Scatter plots of
microwave fluxes, HXR light curves, and red shifts dependence on the
blue shifts of \fexxi~1354.09~{\AA} during the flare precursor. The
correlation coefficients (cc.) are labeled. \label{corr}}
\end{figure}

To investigate the driven-mechanism of this chromospheric
evaporation before the {\it GOES} flare oneset, we firstly choose 9
points from the blue shifts of \fexxi~1354.09~{\AA} during the flare
precursor, i.e., between 00:03:30~UT and 00:05:50~UT, as shown by
the blue crosses in Figure~\ref{corr}~(a). The error bars represent
the uncertainty of the Doppler velocity from the multi-Gaussian
fitting \citep[see also.,][]{Li15}. Secondly, we select 9 nearby
points from microwave (34 GHz: purple, 17 GHz: black) and HXR
(25$-$50~keV: green, 1.0$-$8.0~{\AA} derivative: orange) fluxes.
{\it NoRH}, {\it GOES}, {\it RHESSI} and {\it IRIS} have the time
cadences of 1~s, 2.05~s, 4~s and 16.2~s, respectively, which make
them impossible to correlate one-by-one. Therefore, we use the same
points with the closest time. Figure~\ref{corr}~(a) shows that these
points from different light curves are well correlated during the
flare precursor, indicating that this chromospheric evaporation may
be driven by the nonthermal electrons. Figure~\ref{corr}~(b) shows
the blue shifts of \fexxi~1354.09~{\AA} dependence on microwave
emissions in the frequencies of 17~GHz (diamond) and 34~GHz
(square), and also HXR emissions in 25$-$50~keV (triangle) and
1.0$-$8.0~{\AA} derivative (circle) during the flare precursor,
i.e., between 00:03:30~UT and 00:05:50~UT. As expected from the
electron-driven model of chromospheric evaporation
\citep[see.,][]{Tian15,Li15,Li17}, a high correlation between the
blue shifts of \fexxi~1354.09~{\AA} and nonthermal (microwave or
HXR) emissions is found. For example, the correlation coefficients
(cc.) of 0.97/0.88 are detected between the \fexxi~1354.09~{\AA}
blue shifts and the microwave 17/34~GHz emissions, a correlation
coefficient of 0.87 is observed between the \fexxi~1354.09~{\AA}
blue shifts and the HXR 25$-$50~keV emission, and a correlation
coefficient of 0.88 is found between the \fexxi~1354.09~{\AA} blue
shifts and the SXR 1.0$-$8.0~{\AA} flux derivative. Such high
correlation coefficients demonstrate that the electron beams which
might be accelerated by magnetic reconnection
\citep[e.g.,][]{Kundu94,Brosius07,White03,Asai13} drive the
explosive chromospheric evaporation during the flare precursor
\citep{Tian15,Brosius16,Li17}. Meanwhile, the microwave and HXR
emissions observed by NoRH and {\it RHESSI} images exhibit a
pronounced brightening source that is co-spatial with the
ribbon-like transient during the flare precursor peak, which gives
an additional evidence of electron-driven evaporation before the
{\it GOES} flare onset \citep{Veronig10,Zhang16b,Li17b}. On the
other hand, we also plot the dependence of the blue shifts of
\fexxi~1354.09~{\AA} on the red shifts of \siiv~1402.77~{\AA} during
the flare precursor, and obtain a high correlation coefficient of
0.84. This is consistent with previous findings during solar flare
\citep{Li15,Tian15,Li17,Li17b}, due to the fact that the red shifts
of cool emission line are caused by an over-pressure of the
evaporated material, and should therefore exhibit a correlation with
the blue shifts of hot emission line. We note that such a high
correlation coefficient is not found between the Doppler shifts of
\ci~1354.29~{\AA} and \fexxi~1354.09~{\AA}, which maybe because that
the \ci~1354.29~{\AA} line is formed in a deep layer and not
significantly affected by the condensation plasma during the flare
precursor.

\begin{figure}
\epsscale{1.0} \plotone{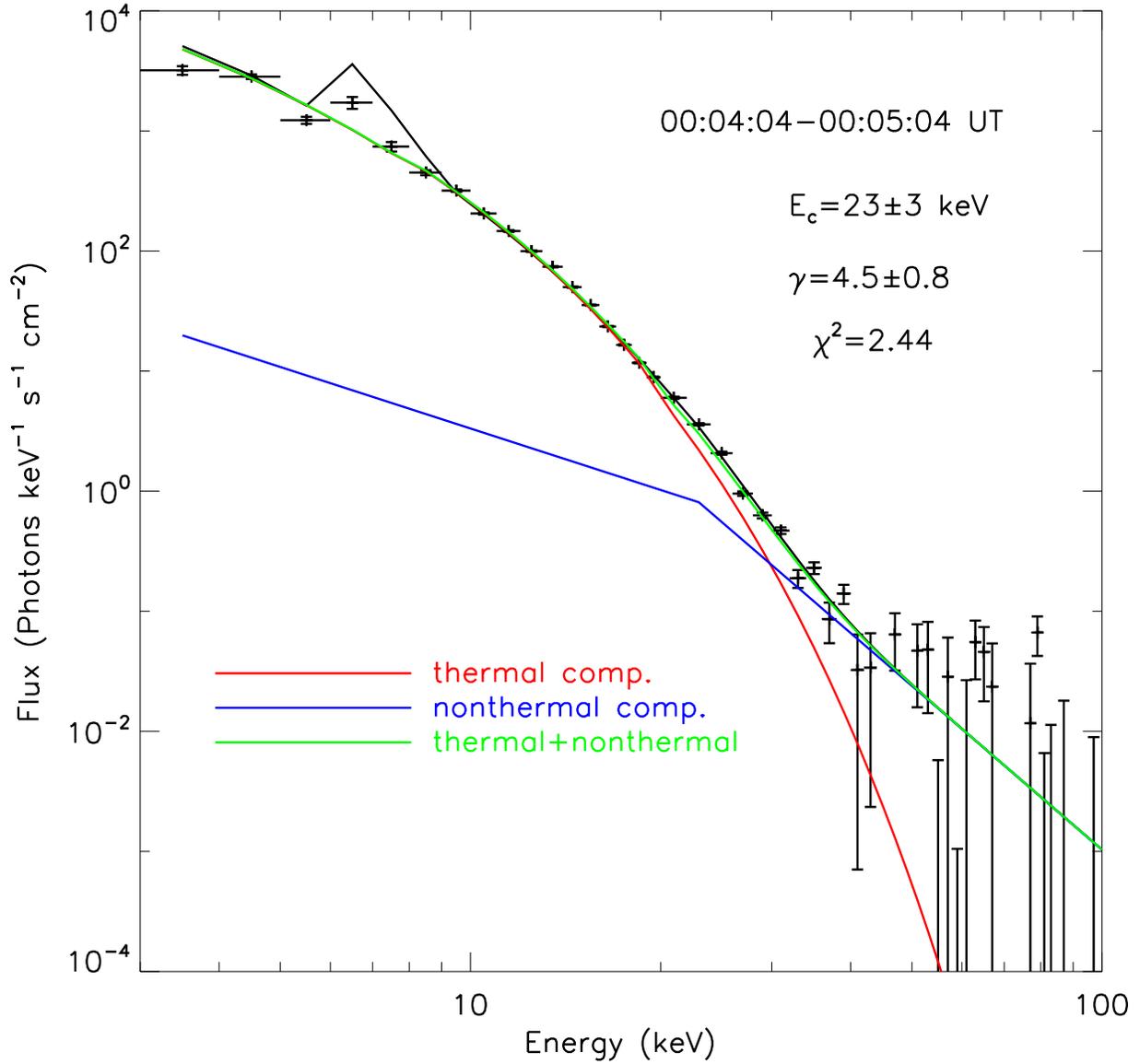} \caption{{\it RHESSI} X-ray spectra
with error bars (black lines) and their fitting results (color
lines) between 00:04:04$-$00:05:04~UT on 2014 October 27. The
spectra for the thermal (vth) component and nonthermal (bpow)
component are shown with the red and blue lines, respectively. The
sum of both components is displayed with the green line. The
integration time and fitted parameters are also given.
\label{hessi}}
\end{figure}

To further understand the deposited energy flux of the precursor
event during chromospheric evaporation, the X-ray spectra observed
by {\it RHESSI} are also analyzed here. Figure~\ref{hessi} shows the
X-ray spectra with error bars and their two-component (thermal and
nonthermal) fitting results  during the flare precursor, i.e., from
00:04:04 to 00:05:04~UT. The physical parameters such as the break
cutoff energy(E$_c \approx$~23$\pm$3~keV), the power-law index
($\gamma \approx$~4.5$\pm$0.8) are derived from the spectral fitting
as well as the Chi-square ($\chi^{2}=2.44$). The Chi-square shows
quite reasonable fitting with $\chi^{2}<3$ presented by
\cite{Sadykov15}. Then we can estimate the total nonthermal power
(P$_{tot}$) of the accelerated electrons from Equation~\ref{ptot}
\citep{Aschwanden05,Zhang16b}.
\begin{equation}
 P_{tot} = 1.16 \times 10^{24} \gamma^{3} I_1 (\frac{E_c}{E_1})^{-(\gamma -1)},
 \label{ptot}
\end{equation}
where I$_1$ represents the photon count rates at energies of
E$\geq$E$_c$, and E$_1$ denotes the lower cutoff energy. In our
observations, $I_1~=~\int_{E_c}^{\infty} I(E)dE~\approx$
5.3$\times$10$^{2}$~photon~s$^{-1}$~cm$^{-2}$ \citep{Aschwanden05}.
Assuming that E$_c$ = E$_1$ \citep{Aschwanden05,Zhang16b}, P$_{tot}$
is estimated to be
$\sim$(5.6$\pm$2.9)$\times$10$^{28}$~erg~s$^{-1}$.
Figure~\ref{image}~(a) shows the HXR sources with the blue contours
set in 70\% and 90\% of the local maximum at 25$-$50~keV. The HXR
source areas are inside these two blue contours, and the values are
estimated in the range of 2.2$\times$(10$^{17}-$10$^{18}$)~cm$^2$.
The projection effect is also considered here
\citep[e.g.,][]{Sadykov15}. Finally, the total nonthermal energy
flux ($P_{tot}/A$) is estimated to be about
(2.5$\pm$1.3)$\times$(10$^{10}-$10$^{11})$~erg~s$^{-1}$~cm$^{-2}$,
which is larger than the typical threshold of
$\sim$10$^{10}$~erg~s$^{-1}$~cm$^{-2}$ for the impulsive evaporation
\citep{Fisher85,Fisher85b,Zarro88}. The received upper nonthermal
energy flux derived from this precursor event is high as much as
that of an X1 flare i.e.,
$\sim$3.5$\times$10$^{11}$~erg~s$^{-1}$~cm$^{-2}$ \citep{Kleint16}.
However, a much higher energy flux
($\sim$1.5$\times$10$^{12}$~erg~s$^{-1}$~cm$^{-2}$) for another
solar flare is reported \citep[see][]{Sharykin17}. Our result
suggests a strong energy flux during the flare precursor. Meanwhile,
the characteristic timescale could be estimated from the nonthermal
pulse in the HXR and microwave emissions, which is about 60~s
(Figures~\ref{flux} and \ref{hessi}). It is short and in the order
of the typical timescale of explosive chromospheric evaporation
\citep{Zarro88,Sadykov15}. All these observational results further
confirm an explosive chromospheric evaporation during the flare
precursor.

\section{Conclusions and Discussions}
Based on the spectroscopic and imaging observations from {\it IRIS},
{\it SDO}, NoRH, {\it RHESSI}, and {\it GOES} we investigate the
temporal and spatial relationships between the blue shifts of
\fexxi~1354.09~{\AA} and the nothermal emissions in microwave
17/34~GHz and HXR 25$-$50~keV before a {\it GOES} M7.1 flare on 2014
October 27. First, a small but well-developed peak in SXR and EUV
pasbands before the {\it GOES} flare onset is identified as a flare
precursor. Second, the hot \fexxi~1354.09~{\AA} line exhibits blue
shifts and the cool \siiv~1402.77~{\AA} line shows red shifts during
the flare precursor, the blue shifts and red shifts are correlated
well with each other. Moreover, the total nonthermal energy flux
during the flare percussor exceeds the critical value
\citep{Fisher85,Fisher85b}, and it is characterized by a short
timescale. All these facts suggest that an explosive chromospheric
evaporation occurs during this flare precursor. Third, the blue
shifts of \fexxi~1354.09~{\AA} show a good correlation with the
microwave/HXR emissions, implying that the explosive chromospheric
evaporation is most likely driven by nonthermal electrons, although
we can not exclude a possible contribution from the heating of
Alfv\'{e}n waves \citep{Reep16,Lee17}.

Although chromospheric evaporation has been investigated in a large
number of studies before
\citep[e.g.,][]{Ding96,Brosius04,Milligan06,Chen10,Zhang13,Tian14,Brosius15,Lee17,Li17c},
to the best of our knowledge, this is the first report of an
electron-driven explosive evaporation during the SXR/EUV precursor
before the {\it GOES} flare onset. We note that chromospheric
evaporation has been detected during the EUV `precursor' peak by
\cite{Brosius04} and \cite{Brosius07,Brosius10}. However, those EUV
`precursors' actually showed up after the {\it GOES} flare onset, or
during the solar flare. In addition, the locations of those EUV
`precursors' seem to be remote from the nonthermal emissions due to
the limited observations, such as HXR sources
\citep{Brosius07,Brosius10}. It is widely accepted that percussor
peak is an important phenomenon prior to a solar flare
\citep{Cheng16a,Bamba17,Li17c,Shen17,Zhang17}. Therefore, the
observational results presented here help us to better understand
the initiation process of solar flares. Electron-driven evaporation
is detected before the flare onset, which implies that magnetic
reconnection has occurred and accelerated the electrons prior to
solar flare. This is also supported by some other observations
\citep[e.g.,][]{Bamba13,Li16b,Bamba17,Li17b,Li17c,Shen17}. The
pre-flare reconnection \citep[usually weak, see][]{Li17c} may cause
the strong magnetic reconnection and trigger the associated solar
flare.

Although the explosive chromospheric evaporation is observed before
the {\it GOES} flare onset, its properties are similar to the
explosive evaporation occurring during solar flare in general.
First, both of the evaporations show a similar temporal and spatial
correlation between the blue shifts (or upflows)/red shifts (or
downflows) of emission lines and the HXR or microwave fluxes/sources
\citep{Milligan09,Veronig10,Tian15,Brosius16,Zhang16a,Lee17,Li17}.
Second, the pre-flare evaporation here also tends to appear at the
front of the ribbon-like transient, which agrees with previous
findings that chromospheric evaporation appears at the outside of
flare ribbon \citep{Czaykowska99,Li04,Li15,Tian15}. Our observations
indicate that the chromospheric evaporation either in the pre-flare
phase or during the flare occurs in successively formed flare loops.
There are also some tiny differences between the explosive
chromospheric evaporation during the flare precursor and that in the
impulsive phase of solar flare
\citep[e.g.,][]{Tian14,Tian15,Li15b,Brosius16}. For example, the red
shifts of the cool \ci~1354.29~{\AA} line appear much more constant
and flat during the flare percussor. They do not show pronounced
precursor peak which corresponds with the blueshifted peak in the
hot \fexxi~1354.09~{\AA} line, this is usually not the case for the
explosive chromospheric evaporation during the impulsive of solar
flare \citep[see.,][]{Li15}.

The maximum speed of the \fexxi~1354.09 line during the flare
precursor is only $\sim$60~km~s$^{-1}$, which is less than previous
findings during the flare impulsive phase
\citep{Tian14,Tian15,Brosius15,Li15,Li15b,Young15,Lee17}. The small
speed might be caused by the projection effect, since the precursor
event occurred somewhat away from the solar disc center, i.e.
$\sim$S10W40. In this case, a projection effect would be involved,
which may affect the estimation of the local plasma velocity from
Doppler shift. But it just affects the value of Doppler velocity,
but does not change its direction (the nature of flows). In previous
observations, the blueshifted speed of an explosive chromospheric
evaporation during a solar flare was often larger than
100~km~s$^{-1}$ \citep{Zarro88,Sadykov15,Kleint16}. However, the
observed blue shift of \fexxi~1354.09~{\AA} is the lower limit of
the local plasma upflow, the actual velocity should be larger. And
we also detect a pronounced redshifted pulse from the cool emission
line in \siiv~1402.77~{\AA}, indicating the local plasma downflow.
This is similar as the simulations of the explosive heating model
\citep{Kostiuk75}, which presented that a temperature region divides
the solar atmosphere into redshifted and blueshifted parts
\citep[e.g.,][]{Livshits83,Kosovichev86,Liu09}, and the division
temperature is $\sim$1~MK. Moreover, the total nonthermal energy
flux during the flare percussor is high enough, and its
characteristic timescale is very short. All these observational
facts are well consistent with the explosive evaporation model
\citep{Fisher85,Fisher85b,Zarro88,Reep15,Kleint16}. In a word, the
projection effect does not change our main results
\citep{Sadykov15}. Our observations also indicate that the speed of
evaporated materials during the flare precursor is possibly smaller
than that during the flare impulsive phase.

\acknowledgments The authors would like to thank the referee for
his/her inspiring and valuable comments, which help us to improve
the manuscript significantly. We also acknowledge Prof. H.~S.~Ji,
Y.~Su, Y.~N.~Su, Q.~M.~Zhang, H.~Tian, Y.~D.~Shen, and X.~L.~Yan for
their fruitful discussions. We thank the teams of {\it IRIS}, NoRH,
{\it RHESSI}, {\it GOES}, {\it SDO} for their open data use policy.
This study is supported by NSFC under grants 11603077, 11573072,
11773079, 11773061, 11403011, 11473071, 11333009, the CRP
(KLSA201708), the Youth Fund of Jiangsu Nos. BK20161095, BK20171108,
XDA15052200, as well as  National Natural Science Foundation of
China (U1731241), `Strategic Pilot Projects in Space Science' of
CAS, (XDA15052200). Dong~Li supported by the Specialized Research
Fund for State Key Laboratories. Ying~Li is also supported by CAS
Pioneer Hundred Talents Program. The Laboratory No. 2010DP173032.

\end{document}